\documentclass{desyproc}

\newcommand{\fbarn} {\ensuremath{\mathrm{fb^{-1}}}}
\newcommand{\ppbar} {\ensuremath{p\overline{p}}}

\newcommand{\pt}{\ensuremath{p_T}}
\newcommand{\et}{\ensuremath{E_T}}

\newcommand{\gev}{\ensuremath{\rm GeV}}
\newcommand{\gevc}{\ensuremath{\rm GeV/c}}
\newcommand{\gevcsq}{\ensuremath{{\rm GeV}/{\rm c}^{2}}}

\begin{document}
%------------------------------------
\title{Review of Recent Tevatron Jet, $W/Z$+Jet and Heavy-flavor Production 
Results}

%for single authors the superscripts are optional
\author{{\slshape Shin-Shan Yu\\\\On behalf of the CDF and D\O\ Collaborations}\\[1ex]
Fermi National Accelerator Laboratory, P.O.~Box~500, Batavia, IL~60510, USA}

% if the proceedings are available online (e.g. at Indico)
% please enter the contribution ID or file_name below for the DOI
\contribID{28}
%\contribID{smith\_joe}

% TO THE CONFERENCE EDITORS: 
% please update the following information      
% before sending the template to the authors
\confID{1407}  % if the conference is on Indico uncomment this line
\desyproc{DESY-PROC-2009-03}
\acronym{PHOTON09} % if you want the Acronym in the page footer uncomment this line
\doi  % if there is an online version we will register DOIs

\maketitle

\begin{abstract} 
This paper reviews several recent measurements at the Fermilab Tevatron,
 including cross sections for inclusive jet, dijet production, cross
 sections for electroweak boson ($W$ or $Z$) production in association with 
inclusive or heavy-flavor ($b$ or $c$) jets, and $b$-jet shapes. 
In addition, searches for new physics using the dijet angular distributions
 are discussed. 
These analyses are based on integrated luminosities of 0.3--2.5 \fbarn\ of 
\ppbar\ collisions at $\sqrt{s}=1.96$~TeV, collected with the CDF and D\O\ 
detectors. The results directly test the leading order and next-to leading 
order calculations of perturbative quantum chromodynamics and provide 
constraints on the parton distribution functions and physics beyond the 
standard model.
 
\end{abstract}

\section{Introduction \label{sec:intro}}
Measurements using jet final states have been of great interest to both 
experimentalists and theorists for the following reasons. First, 
among high $\pt$ physics processes at a hadron collider, jet production 
has the largest cross section. Therefore, jet 
production can test perturbative quantum chromodynamics (pQCD) with the 
highest reach in energy and rapidity ($y$). Second, measurements at the 
Tevatron, which are complementary to the measurements by HERA and fixed 
target experiments, may constrain parton distribution functions (PDFs) in the 
region of large $Q^2$ and medium-to-large $x$ and reduce uncertainties on the 
gluon, $b$, and $s$ quark PDFs. Measurements will have greater impacts on PDFs 
when the uncertainties on the cross sections due to variation of 
renormalization and factorization scales (pQCD uncertainties) are much smaller 
compared to the uncertainties from existing PDFs, {\it e.g.} 
measurements of inclusive jet cross section. 
Third, these measurements not only provide stringent 
tests of the standard model (SM) physics, but also probe physics beyond the SM.
 The production of $W$ or $Z$ in conjunction with inclusive or heavy-flavor 
jets is one of the major backgrounds to searches for SM Higgs, SUSY, 
and other models. Measurements of the cross sections of these processes 
decrease uncertainties on the estimation of backgrounds. 
The angular distributions of jet events, which are not very sensitive to 
PDFs, can also probe the presence of new physics. 

Section~\ref{sec:def} briefly describes the jet definition and 
reconstruction algorithms used at the Tevatron. 
Sections~\ref{sec:incj}--\ref{sec:bjetshape} discuss the results of these 
analyses. Section~\ref{sec:con} gives the conclusion.

\section{Jet Definition and Reconstruction \label{sec:def}}
Jets are collimated sprays of particles originating from quarks or gluons. 
The most common jet reconstruction algorithms at the Tevatron are 
midpoint cone and $k_T$.\footnote{When comparing data and theory, the same 
algorithms are applied.} The midpoint cone and $k_T$ algorithms cluster 
objects\footnote{In data, the ``object'' is a calorimeter cell 
with energy deposit. In theory, the ``object'' is a parton.} based 
on their proximity in the geometry and momentum space, respectively. 
The midpoint cone algorithm starts from objects above an energy threshold 
(seeds) and sums the four-momentum vectors of all objects 
within a cone of radius $R_\mathrm{cone}$\footnote{$R_\mathrm{cone}^2\equiv \Delta y^2 + \Delta \phi^2$.} around the seed. 
The total four-momentum vector of these objects defines a new jet axis. 
The process is iterated until the updated jet axis is within a tolerance from 
the previous jet axis; a stable cone is formed. Then, additional seeds 
are added at the midpoints between all pairs of stable cones whose separation 
is less than $2R_\mathrm{cone}$ and the clustering procedure is repeated 
using these additional seeds. Finally, geometrically overlapping cones are 
split or merged depending on the amount of shared momentum. The $k_T$ 
algorithm starts by considering every object as a protojet and calculates 
$k_{T,i}^2$ for each protojet and $k_{T(i,j)}^2$ for each pair of 
protojets.\footnote{Here, $k_{T,i}^2\equiv p_{T,i}^2$ and 
$k_{T(i,j)}^2 \equiv min(p_{T,i}^2,p_{T,j}^2)\Delta R_{i,j}^2/D^2$, where 
$R_{i,j}$ is the distance between the two protojets in the $y-\phi$ space 
and $D$ is a parameter that controls the size of the jet.} 
All $k_{T,i}^2$ and $k_{T(i,j)}^2$ are then collected into a single sorting 
list. If the smallest in this list is $k_{T,i}^2$, protojet $i$ is 
promoted to a jet and removed from the list. If the smallest is 
 $k_{T(i,j)}^2$, protojets $i$ and $j$ are combined into a single protojet. 
The procedure is iterated until the list is empty. 
The cone algorithm has simpler underlying event and multiple interaction 
corrections while the $k_T$ algorithm is less sensitive to higher order 
perturbative QCD effects. 
More discussions of the strengths and weaknesses of these two algorithms 
 are in Ref.~\cite{Ellis:2007ib}.

Three levels of energies are defined, (i) parton level: the true 
energy of the parent parton (quark or gluon), (ii) particle level: the total 
true energy of all particles contained in a jet, including underlying event 
and products of fragmentation and hadronization, but excluding the energy 
from multiple $p\bar{p}$ interactions per crossing, (iii) detector level: 
energy measured in the calorimeters. 
The cross sections discussed here are presented as functions of 
particle-level energy.\footnote{The energy at the particle level depends only 
on physics models, not detectors.} Calorimeters may 
under- or over-measure the energies of particles due to finite resolution, 
non-uniformity, and inefficiency of detector. Programs that provide 
theoretical predictions of cross sections at the next-to leading order (NLO) 
typically do not include parton showering. 
Therefore, in order to have a valid comparison between data 
and theory, corrections have to be applied. For measurements in data, 
corrections of energy from the detector to the particle level follow the 
procedures described in Ref.~\cite{Bhatti:2005ai,D0jes}.\footnote{The 
corrections are $\approx 20$\% (50\%) of the jet energy at 50~GeV and 
$\approx 10$ \% (20\%) at 400~GeV for CDF(D\O).} 
For theory predictions, corrections of energy from the parton to the 
particle level are obtained by comparing {\tt PYTHIA} or {\tt HERWIG} MC with 
parton shower and fragmentation switched on vs. switched off.\footnote{The 
corrections are $\approx 10$--20\%\ at 50~GeV and drops quickly to below 5 
\%\ when energy is above 100~GeV.}

\section{Measurements of Inclusive Jet Cross Section \label{sec:incj}}
As mentioned in Section~\ref{sec:intro}, inclusive jet production cross section
 provides constraints on the gluon PDF.\footnote{The inclusive jet cross 
section measured in the forward region will be most sensitive to gluon PDF 
since new physics is expected to appear mostly in the central region.} 
The inclusive jet cross section from Tevatron Run~I~\cite{Abe:1996wy} had 
excess in data with respect to NLO predictions at high \pt. 
%Variation of different PDFs at the time ($\sim$10 years ago) was not enough 
%to cover the data and theory difference. 
Data had been included later in the global fits of CTEQ6 and 
MRST2001 and preferred larger contribution of gluons at high $x$.
At Run II, CDF and D\O\ have measured inclusive jet 
cross section with midpoint cone~\cite{Abulencia:2005yg,:2008hua} 
and $k_T$ algorithms~\cite{Abulencia:2005jw}. The Run II measurements 
have extended the cross section reach significantly both in \pt\ and 
rapidity ($y$). The midpoint seeds are added\footnote{There were 
no midpoint seeds at Run I.} in the cone algorithm 
in order to reduce sensitivity to non-perturbative effects, such 
as radiation of soft gluons. 
%A cone size of $R_\mathrm{cone}=0.7$ and a $k_T$ size control parameter 
%$D=0.7$ are used to reconstruct jets, where the total theory correction 
%for out-of-cone energy and underlying event is minimal. 

\begin{figure}[hb]
\centerline{\includegraphics[width=0.65\textwidth]{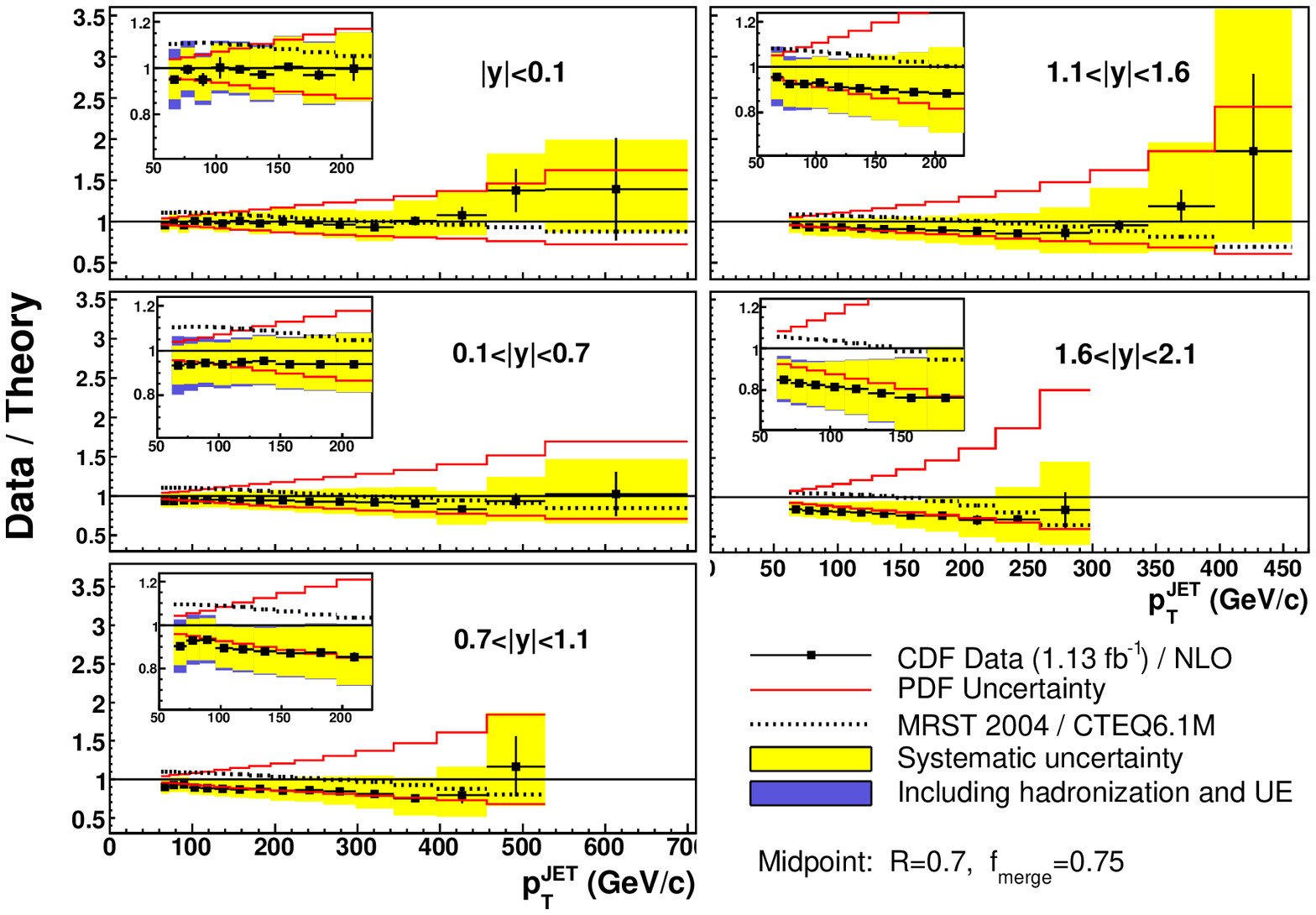}}
\centerline{\includegraphics[width=0.65\textwidth]{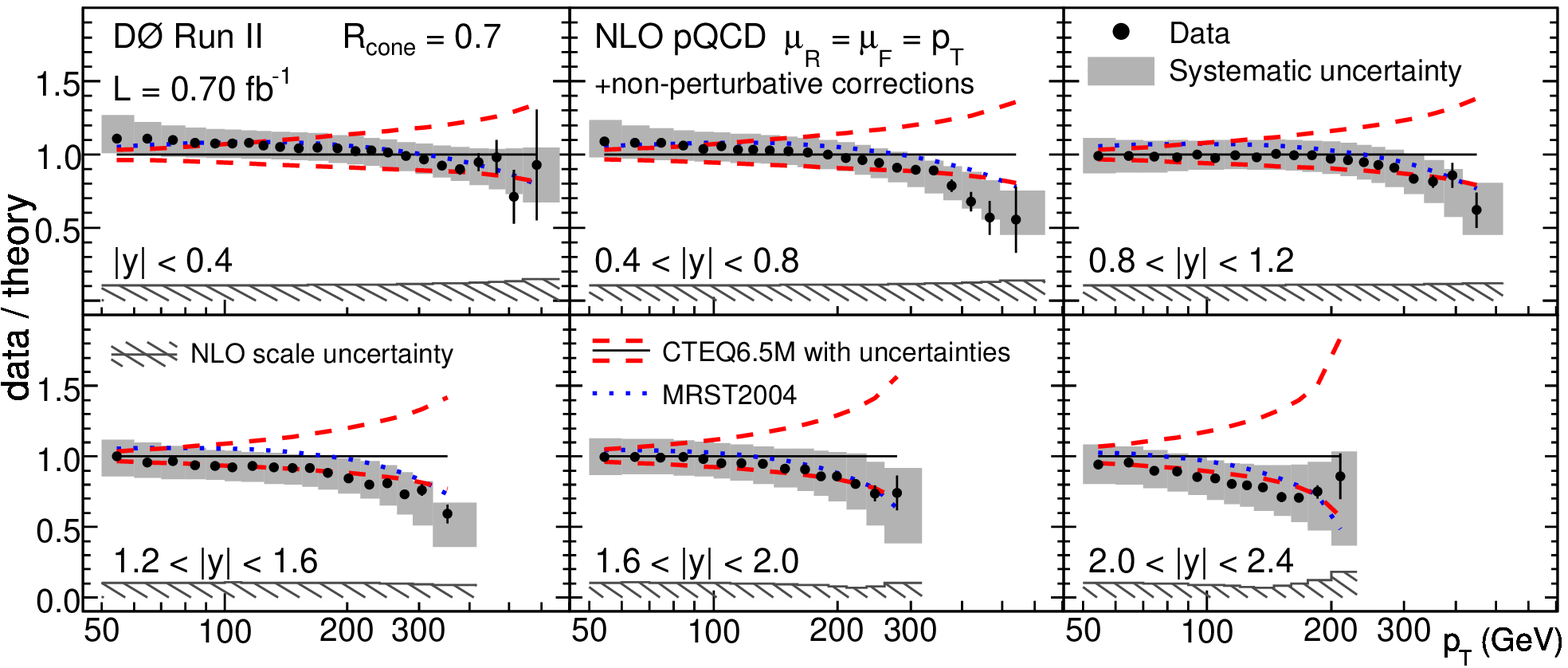}}
\caption{The Tevatron Run II results of inclusive jet 
 cross section using midpoint cone algorithm. Ratios of 
 CDF (top) and D\O\ (bottom) data to NLO theory are shown.}\label{fig:incj1}
%\end{figure}

%\begin{figure}[hb]
\centerline{\includegraphics[width=0.65\textwidth]{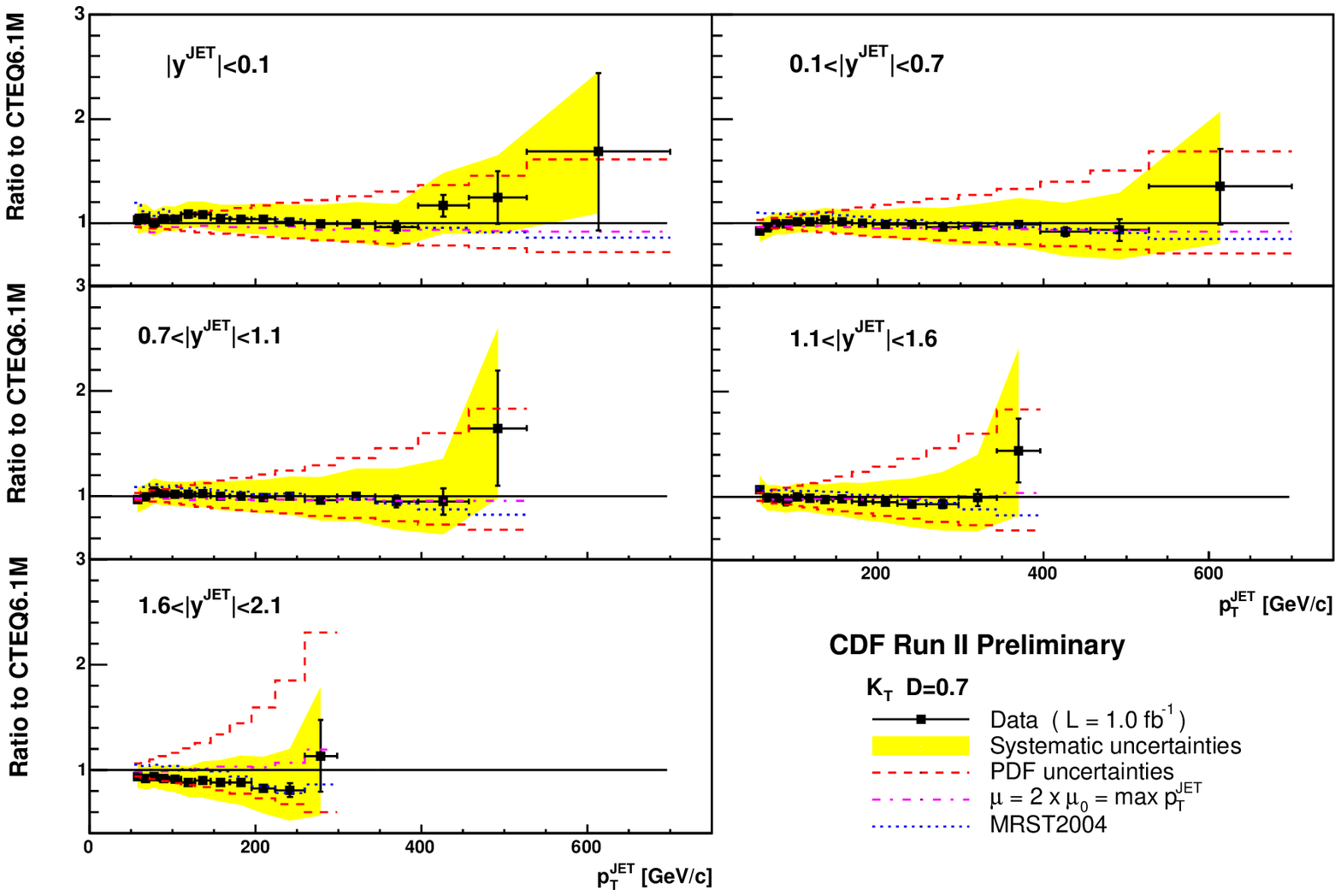}}
\caption{The Tevatron Run II results of inclusive jet 
 cross section using $k_T$ algorithm. Ratios of 
 CDF data to NLO theory are shown.}\label{fig:incj2}
\end{figure}

The cross section is measured as a 
function of corrected 
jet \pt\ (to the particle level), in 5--6 bins of jet rapidity. 
Dominant sources of systematic uncertainties are jet energy scale\footnote{The uncertainty on the jet energy scale is 2--3(1.2--2)\%\ for 
CDF(D\O).} and jet energy resolution. 
Measurements in data are compared to NLO predictions and CTEQ6.1M PDFs for 
CDF, CTEQ6.5M PDFs for D\O. The renormalization and factorization 
scales ($\mu_R$ and $\mu_F$) are set to $0.5\pt^\mathrm{jet}$ for CDF and 
$\pt^\mathrm{jet}$ for D\O. Figure~\ref{fig:incj1} and 
Figure~\ref{fig:incj2} show the ratios of Run II data to theory using the 
cone algorithm and $k_T$ algorithm, respectively. Although the PDFs and 
scales used are not exactly the same, all three measurements have a 
similar trend: at high \pt\ and large $|y|$ (equivalent to large $x$),
 the data prefer smaller values of cross section than the theory prediction. 
The CDF $k_T$ and D\O\ cone measurements are already included 
in the global fit of MSTW2008 PDFs; not only the uncertainties of gluon 
component have decreased, but also the central values. 
There is an ongoing effort to include the CDF cone measurement and update
 CTEQ PDFs as well.

\section{Measurements of Dijet Mass Spectra and Search for New Particles 
Decaying into Dijets \label{sec:dijetmass}}
Measurements of dijet mass spectra provide an alternate method to 
constrain PDFs. In addition, new particles predicted by physics 
beyond SM may appear as resonances in the dijet mass spectra. These 
new particles and decays include: (i) $q^*\rightarrow qg$ (quark 
compositeness~\cite{excitedq}), (ii) axigluon or coloron $\rightarrow q\bar{q}$ (chiral color model~\cite{color}), (iii) color-octet techni-$\rho$ ($\rho_{T8}$) $\rightarrow q\bar{q}$ or $gg$ (extended and topcolor-assisted technicolor~\cite{technicolor}), (iv) Randall Sundrum graviton $\rightarrow q\bar{q}$ or $gg$ (warped extra dimension~\cite{RS}), (v) $W^{\prime}$ ($Z^{\prime}$) $\rightarrow q\bar{q}^{\prime} 
(q\bar{q})$ (grand unified theories GUT~\cite{GUT}), (vi) diquark $\rightarrow 
qq$ or $\bar{q}\bar{q}$ ($E_6$ GUT~\cite{E6}). 
The CDF measurement of dijet mass spectrum~\cite{Aaltonen:2008dn} 
requires both jets to be central ($|y^\mathrm{jet}| < 1.0$) while the D\O\ 
measurement~\cite{D0dijetmass} is performed in six bins of $|y|$ and 
extended to $|y|_\mathrm{max}=2.4$, where $|y|_\mathrm{max}$ is the rapidity 
of the jet with the largest $|y|$ among the two leading jets (see Figure~\ref{fig:dijetmass}).\footnote{Ordered in jet \pt.} Both CDF and 
D\O\ have not seen significant discrepancy from the NLO predictions and the 
results are yet to be included in the global PDF fits. While the limits on 
$W^{\prime}$, $Z^{\prime}$, and RS graviton are not as stringent as those 
obtained by the lepton channels, CDF has set the world's best 
limits and excluded at 95\%\ C.L. the mass of $q^*$ at 260--870~\gevcsq, 
of axigluon and coloron at 260--1250~\gevcsq, of $\rho_{T8}$ at 
260--1100~\gevcsq, and of $E_6$ diquark at 260--630~\gevcsq. The D\O\ 
limits are work in progress.

\section{Search for New Physics in the Dijet Angular Distributions}
\begin{figure}[hb]
\begin{center}
\includegraphics[width=0.4\textwidth]{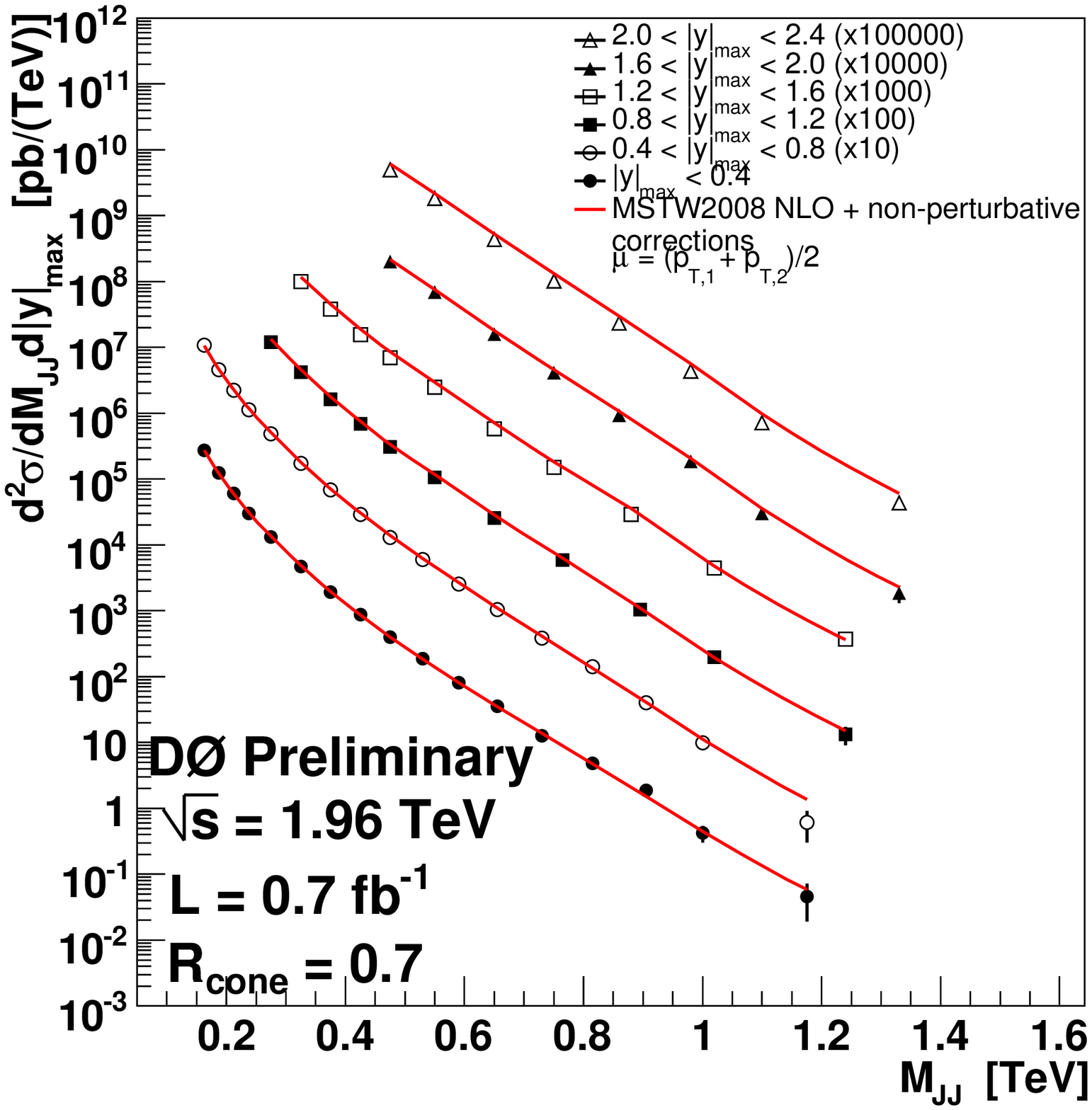}
\includegraphics[width=0.5\textwidth]{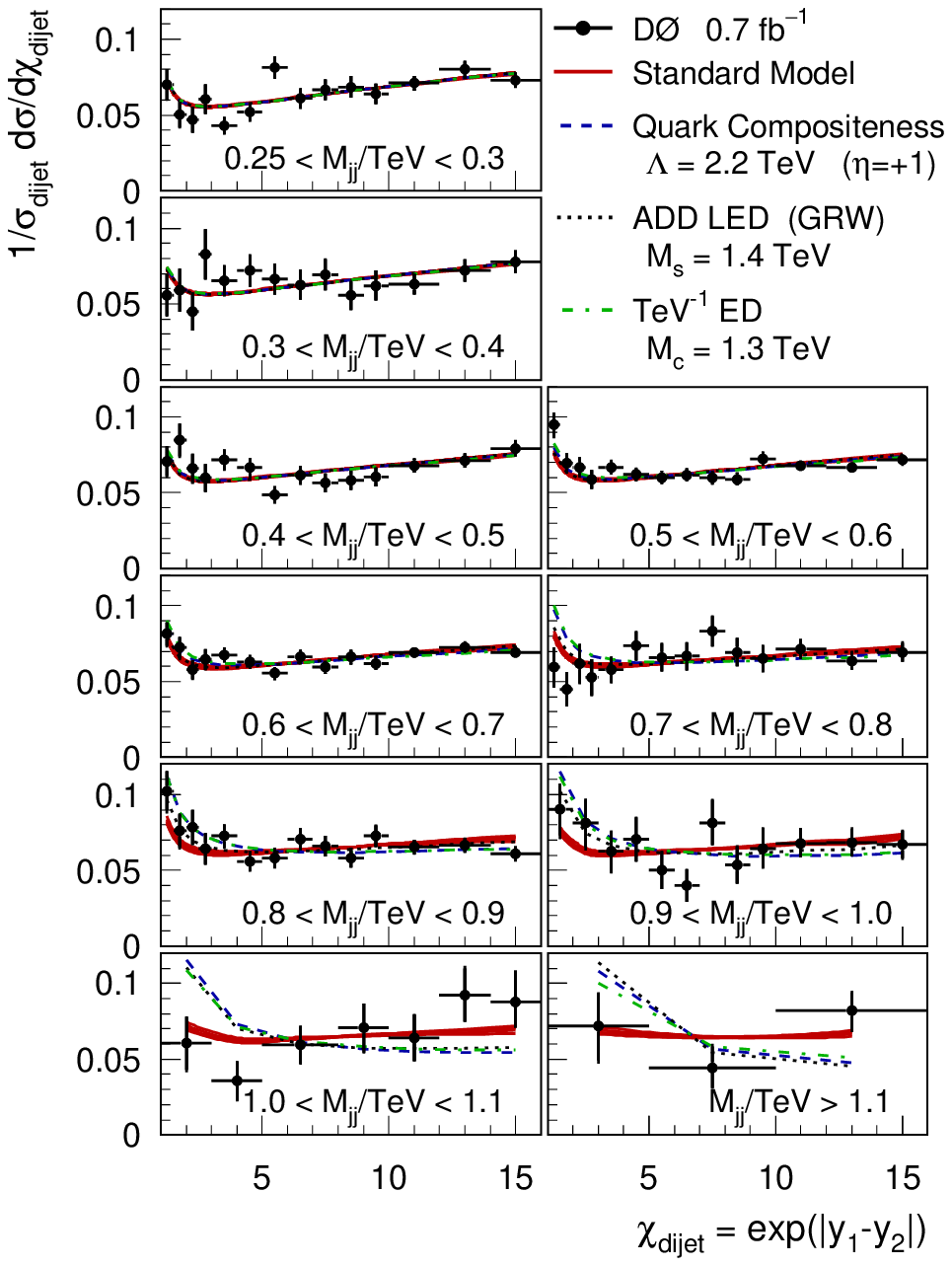}
\end{center}
\caption{Measurements of dijet mass spectra by D\O\ (left) and  
	 normalized $\chi_\mathrm{dijet}$ distributions 
	from the D\O\ data, SM, and new physics predictions (right). 
	}
	\label{fig:dijetmass}
\end{figure}

An excess in data may indicate presence of new physics, but may also 
imply that the PDFs have to be updated; matrix elements for hard 
scattering processes and PDFs are entangled in the calculation of 
absolute production cross sections. Instead, the shapes of 
angular distributions, which are disentangled from PDFs, are more sensitive 
to new physics. The shape of the dijet angular variable, 
$\chi_\mathrm{dijet}$\footnote{Here, $\chi_\mathrm{dijet}\equiv (1+\cos\theta^*)/(1-\cos\theta^*)$, where $\cos\theta^*=\tanh(y^*)$, 
$\pm y^*$ is the rapidity of each jet in the center-of-mass frame, and 
$y^*=\frac{1}{2}(y_1-y_2)$.}, is flat for Rutherford scattering, and is more 
strongly peaked at small value of $\chi_\mathrm{dijet}$ in the presence of 
new physics;\footnote{Here, the new physics models refer to quark compositeness, large extra dimension, and TeV$^{-1}$ extra dimension.} the peak fraction 
increases as the dijet mass $M_{jj}$ increases.
CDF has focused on $M_{jj}=$0.55--0.95~TeV/$c^2$ and looked at the ratio 
of the number of events in two $\chi_\mathrm{dijet}$ regions: 
$N_{1<\chi_\mathrm{dijet}<10}/N_{15<\chi_\mathrm{dijet}<25}$, for four 
$M_{jj}$ bins~\cite{CDFdijetang}. 
D\O\ has a wider mass range\footnote{This is the same dataset that is used to 
measure the dijet mass spectrum, as described in Section~\ref{sec:dijetmass}.},
  0.25--above 1.10~TeV/$c^2$, and has studied the normalized 
$\chi_\mathrm{dijet}$ distributions for ten $M_{jj}$ bins (see 
Figure~\ref{fig:dijetmass})~\cite{Collaboration:2009mh}. Since no 
significant 
discrepancy is observed between the data and SM prediction, both experiments 
set limits on the compositeness scales~\cite{excitedq}, $\Lambda_C$, which 
characterizes the physical size of composite states. D\O\ has obtained the 
world's best limits: $\Lambda_C >$ 2.84 (2.82) TeV for the interference term
 $\eta=+1(-1)$, assuming flat prior in the new physics cross section. 
D\O\ also set limits on ADD large extra dimension~\cite{ADD} and 
TeV$^\mathrm{-1}$ extra dimension~\cite{TeV}.

\section{Measurement of $W$ + Inclusive Jet Cross Section \label{sec:wjet}}
CDF has used early Run~II data and measured the $W$ + jet cross 
section~\cite{Aaltonen:2007ip}. While most jet cross section measurements have 
major uncertainties from the jet energy scale, this measurement also 
suffers from the uncertainty on the background estimate at large jet \pt\ 
and high jet multiplicity; this is the region where top pair production 
dominates. 
Measured results are compared with NLO predictions from MCFM and two 
different schemes of interfacing leading-order (LO) matrix element with 
parton shower generators and jet matching (MLM:{\tt ALPGEN}+{\tt HERWIG}+{\tt MLM}, SMPR: {\tt MADGRAPH}+{\tt PYTHIA}+{\tt CKKW}). 
Both LO and NLO predict well the cross section ratios of different jet 
multiplicity $\sigma_n/\sigma_{n-1}$. The NLO predictions also have 
good agreement with the measurement, both in shape and absolute cross 
section, as functions of jet multiplicity and jet \et. 
%(see Figure~\ref{fig:dijetang}). 
As expected, the LO tends to under-predict the absolute 
cross section. 
Among the two LO schemes, SMPR has better agreement at low \et\ due to 
a better underlying event model in {\tt PYTHIA}.

%\begin{figure}[hb]
%\includegraphics[width=0.5\textwidth]{fig/D0_dijetangular}
%\includegraphics[width=0.5\textwidth]{fig/wjet_3LO}
%\caption{Left: Normalized $\chi_\mathrm{dijet}$ distributions 
%	from the D\O\ data, SM, and new physics predictions. 
%	Right: Ratios of $W$ + jet cross sections measured from CDF data 
%	to LO and NLO predictions, as functions of 1$^\mathrm{st}$ and 
%	$2^\mathrm{nd}$ leading jet \et.}
%	\label{fig:dijetang}
%\end{figure}

\section{Measurements of $W$ + Heavy-flavor Jet Cross Section \label{sec:wbc}}
The production of $W$ boson in association with heavy-flavor jets is one of 
the major backgrounds to searches for new physics ({\it e.g.} Higgs). 
A sample of $W$ boson with heavy-flavor jets may be obtained by requiring 
the jets to contain either secondary vertices ({\tt SECVTX} tagging) or 
a soft electron or muon (soft lepton tagging). 

CDF has measured the $W$ + $b$ jet production cross section, where the 
measurement is proportional to the number of $b$ jets and restricted to 
the kinematic range: a charged lepton with $\pt > 20$~\gevc\ and 
$|\eta| < 1.1$, a neutrino with $\pt > 25$~\gevc, and one or two jets 
regardless of species with $\et > 20$~\gev\ and $|\eta| < 2.0$~\cite{CDFwb}. 
This definition of cross section has been chosen in order to minimize 
uncertainties on the acceptance.
% due to the modeling of production in simulations. 
%Only events with exactly one or two high \et\ jets are selected. 
The jets are tagged by ultra-tight {\tt SECVTX}~\cite{SECVTX}.\footnote{The 
ultra-tight {\tt SECVTX} is operated at a different point from the standard 
{\tt SECVTX}~\cite{SECVTX} and further decreases the light (charm) backgrounds 
by a factor of 10 (4), at the expense 50\%\ reduction in $b$-tagging 
efficiency.} The fraction of tagged jets originating 
from $b$ quarks is extracted by fitting the mass reconstructed at the 
secondary vertices to templates of light, $c$, and $b$-flavor jets 
(see Figure~\ref{fig:wbc}). The cross section has been measured to be 
$2.74\pm 0.27(stat)\pm 0.42(syst)$ pb, which is $\approx 3.5$ 
times larger than the LO prediction of 0.78 pb from {\tt ALPGEN}. 
The NLO calculations are available, but not yet implemented in an MC program 
that allows comparison of data and theory with user-defined 
kinematic requirements.

CDF and D\O\ have also studied samples of $W$ boson with single charm 
candidate by tagging the charm 
quark with soft muon tagging~\cite{:2007dm,Abazov:2008qz}. 
While {\tt SECVTX} and soft lepton taggings could help separating 
heavy-flavor from light-flavor jets, a separation between $b$ and $c$ 
requires more advanced analysis techniques, such as neural network. 
Nevertheless, one could employ the fact that in $W$ + single charm events, 
the muon from semileptonic decays of $c$ hadrons and the charged lepton 
from $W$ decays are oppositely charged, therefore, with a large asymmetry 
in the number of oppositely-charged vs. same-charged events, while 
background from $Wb\bar{b}$ or $Wc\bar{c}$ has zero asymmetry. 
CDF has measured the absolute cross section for $W\rightarrow 
\ell\bar{\nu}_{\ell}$, $\pt^c > 20$~\gevc, and 
$|\eta^c| < 1.5$ to be $9.8\pm 3.2$~pb. D\O\ has measured the cross-section 
ratio, $\sigma(W+c)/\sigma(W+jet)$ for jet \pt\ $> 20$~\gevc\ and $|\eta| < 2.5$, to be $0.074 \pm 0.019 (stat) {+0.012 \atop -0.014 }(syst)$, and also 
measured the ratios as a function of jet \pt. 
%(see Figure~\ref{fig:wbc}). 
Both experiments have found good agreement between data and LO or NLO 
predictions within uncertainties. Since the dominant process 
of $Wc$ production is $gs\rightarrow Wc$, future $Wc$ cross 
section measurements with reduced uncertainties may constrain the $s$ quark 
PDF.

\section{Measurements of $Z$ + Inclusive Jet Cross Section \label{sec:zjet}}
The measurements of $Z$ boson production in association with inclusive jets 
contain only small amount of background from mis-identified leptons 
and are one of the cleanest channels to test pQCD. CDF has measured the $Z$ + 
jet cross section as functions of jet multiplicity and jet \et~\cite{:2007cp}. 
In addition, D\O\ has measured the cross section as a function of $Z$ boson 
kinematics: $\pt(Z)$ and $y(Z)$, and 
the angular separation between $Z$ and jets: $\Delta\phi(Z,\mathrm{jet})$, 
$\Delta y (Z,\mathrm{jet})$, and $y_\mathrm{boost}(Z+\mathrm{jet})$~\cite{Abazov:2008ez,Abazov:2009av,D0zang}. 
%The scales $\mu_R$ and $\mu_F$ have been set to $\mu^2= M_Z^2 + \pt^2(Z)$. 
Both CDF and D\O\ have seen good agreements between data and theory when NLO 
predictions are available. D\O\ has also compared their results with a number of LO matrix element generators and pure parton showering programs, such as 
{\tt ALPGEN}, {\tt SHERPA}, {\tt PYTHIA}, {\tt HERWIG}. Overall, the LO MC 
programs under-predict the cross sections, but the programs that interface 
matrix element generator with parton shower MC have better agreement with 
data in shapes. 
%(see Figure ~\ref{fig:zjet}). 
The results of these comparisons may provide inputs to the MC generation for 
LHC experiments.

%\begin{figure}[hb]
%\begin{center}
%\includegraphics[width=0.4\textwidth]{fig/Pt_12_zjet}
%\includegraphics[width=0.4\textwidth]{fig/D0_Zjetmu3}
%\end{center}
%\caption{Measurements of $Z$ + jet cross section as functions of 
%	first and second leading jet \et\ from CDF data and the ratios of 
%	data to NLO 
%	predictions (left). Comparisons of the cross section as a function 
%	of $| \Delta y(Z,\mathrm{jet})|$, between D\O\ data and the LO 
%	predictions using various MC generators, are also shown (right).}
%	\label{fig:zjet}
%\end{figure}

\section{Measurement of $Z$ + Heavy-flavor Jet Cross Section \label{sec:zb}}
CDF has measured the ratio of $Z$+$b$ jet cross section to 
inclusive $Z$ cross section~\cite{CDFzb}. Measuring the ratio, instead of 
the absolute cross section, makes the systematic uncertainties from luminosity
 and lepton identification largely cancel. Analysis requires at least one jet 
tagged by the standard 
{\tt SECVTX} algorithm and the $b$ fraction is extracted by fitting the 
secondary vertex mass as described in Section~\ref{sec:wbc}. 
The per jet cross section ratio, $\sigma^\mathrm{jet}(Z+b~\mathrm{jet})/\sigma(Z)$, for $\et^{b~\mathrm{jet}}>20$~\gev, $|\eta^{b~\mathrm{jet}}| < 1.5$, 
$76 < M_{\ell\ell} < 106$~\gevcsq, has been measured to 
be $\left(3.32 \pm 0.53 (stat) \pm 0.42 (syst)\right)\times 10^{-3}$. 
Although the measured results are consistent with predictions from MCFM, 
the predictions have a large dependence on scales, which is unexpected 
for NLO calculations. For example, the cross section ratio at 
$N_\mathrm{jet}=2$ for $Q^2 = \left < p_{T,\mathrm{jet}}^2\right>$ is a factor of two of the prediction for $Q^2 = m_Z^2$. 
%(see Figure~\ref{fig:zb}).  
Several investigations show that MCFM does not provide full NLO 
predictions for one of the production diagrams: $q\bar{q}\rightarrow Zb\bar{b}$ when only one $b$ jet is observed.\footnote{When the two $b$ quarks are 
collinear, they may be reconstructed as single $b$ jet. When the two 
$b$ quarks are well separated, one of them may be outside of the 
detector acceptance.} Similar to the case of $W$ + $b$ cross section, 
NLO calculations are available, but not yet implemented in an MC program 
that allows user-defined kinematic requirements. The other dominant 
production process is gluon initiated, $gb\rightarrow Zb$\footnote{Equivalent 
to $gg\rightarrow Zb\bar{b}$.}, therefore, future $Z$ + $b$ 
cross section measurements may constrain the $b$ quark PDF.

\section{Measurement of $b$-jet Shapes \label{sec:bjetshape}}
\begin{figure}[hb]
\begin{center}
\includegraphics[width=0.4\textwidth]{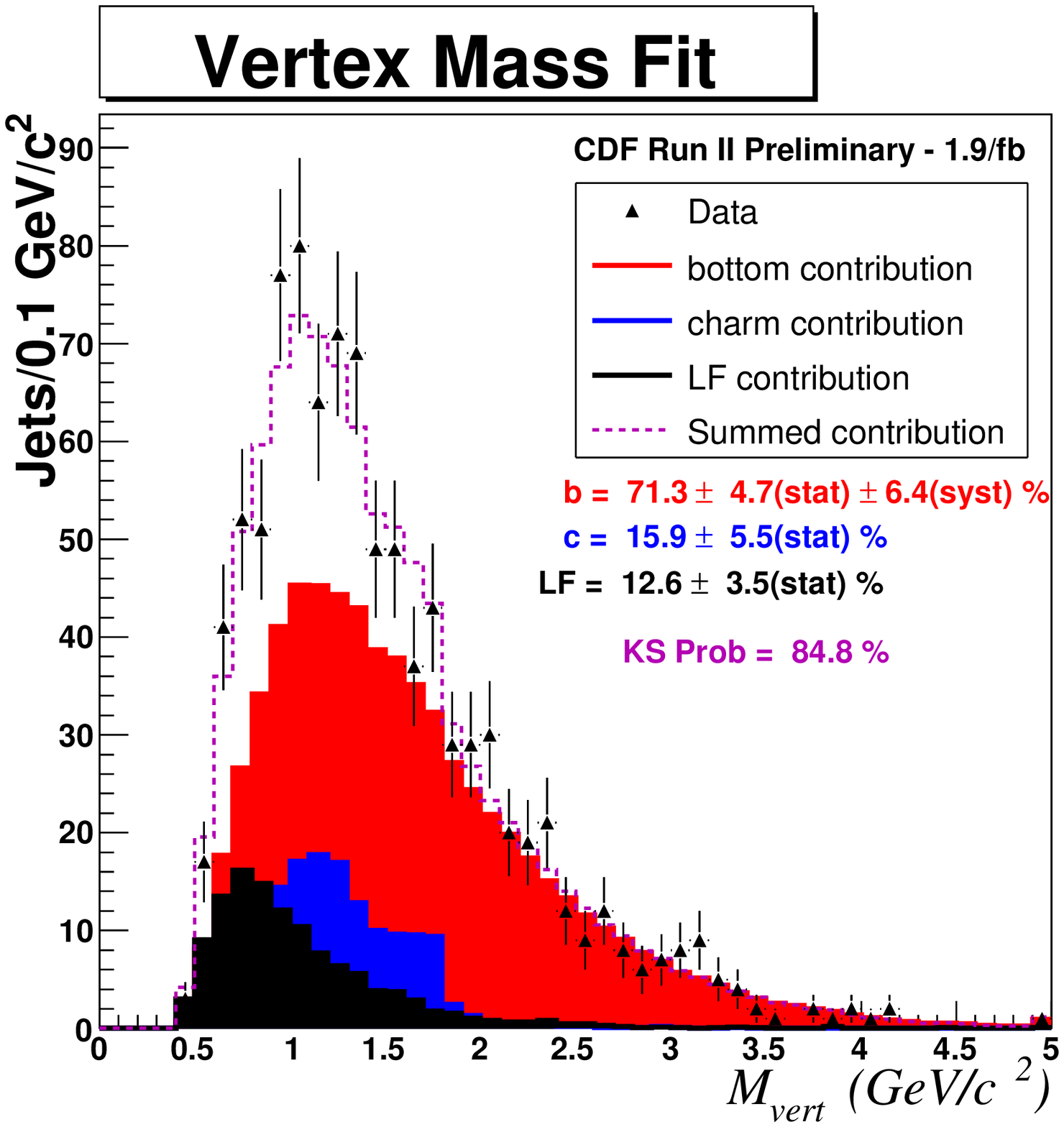}
\includegraphics[width=0.4\textwidth]{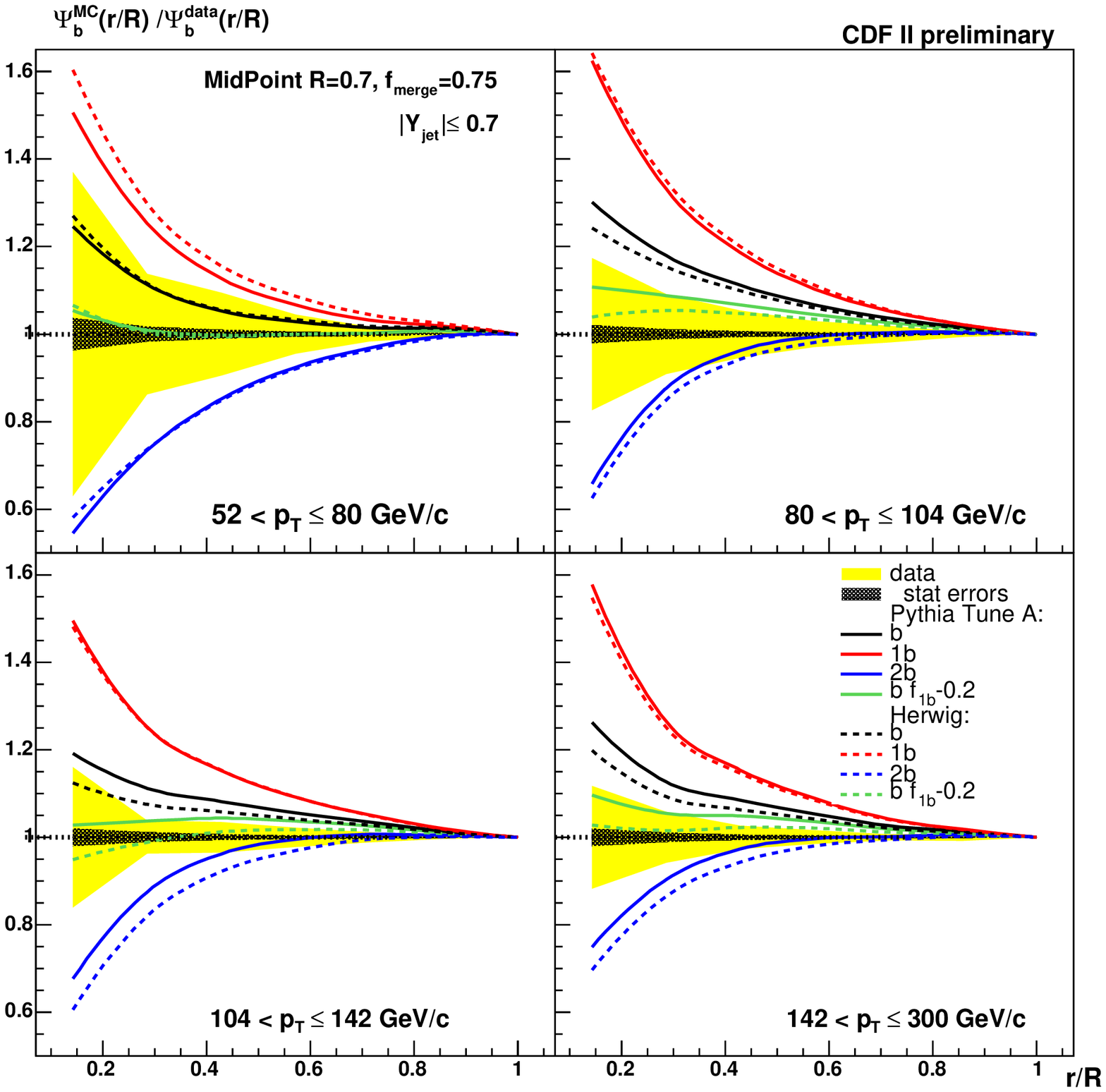}
\end{center}
\caption{Fitting of secondary vertex mass measured in CDF data 
        to templates of light, $c$, and $b$-flavor jets (left). 
%	Right: Ratios of $W$ + $c$ to $W$ + inclusive jet cross sections 
%	measured in D\O\ data, as a function of jet \pt.
	The CDF measurement of $b$-jet shape in four \pt\ bins, and  
	predictions from {\tt PYTHIA} and {\tt HERWIG} with various 
	1-$b$-quark fractions (right).
	}
	\label{fig:wbc}
\end{figure}

The jet shape $\Phi(r)$ is defined as the fraction of momentum carried by 
particles within a cone of radius $r$, relative to the total momentum 
within the jet cone size $R$. 
By definition, $\Phi(R)$ is equal to one. The $b$-jet shapes provide 
an alternate method to probe the $b\bar{b}$ production mechanism, 
particularly the fraction of gluon splitting, which is complementary to 
the measurement of the $b\bar{b}$ angular correlation. 
A $b$-jet that originates from only one $b$ quark has narrower\footnote{
Narrower jet shape means more momentum at small $r$.} jet 
shape than a $b$-jet that originates from two $b$ quarks; gluon splitting 
tends to produce more 2-$b$-quark jets. 
The CDF measurement has been compared to predictions by {\tt PYTHIA} and 
{\tt ALPGEN}, with the default 1-$b$-quark fraction $f_{1b}$, only one $b$ 
quark, only two $b$ quarks, and with $f_{1b}-0.2$~\cite{Aaltonen:2008de}. 
Data have shown a preference over $f_{1b}-0.2$ (see Figure~\ref{fig:wbc}).

%\begin{figure}[hb]
%\begin{center}
%\includegraphics[width=0.4\textwidth]{fig/CDF_zbjet}
%\includegraphics[width=0.4\textwidth]{fig/ratio_unfolded_b_1b_2b}
%\end{center}
%\caption{Left: The CDF measurement of the ratio of $Z$+$b$ jet cross section 
%	relative to inclusive $Z$ cross section, as 
%	functions of $N_\mathrm{jet}$ and $N_{b~\mathrm{jet}}$. 
%	Predictions by MCFM with two different scales are shown. 
%	Right: The CDF measurement of $b$-jet shape in four \pt\ bins, and  
%	predictions from {\tt PYTHIA} and {\tt HERWIG} with various 
%	1-$b$-quark fractions.
%	 }
%	\label{fig:zb}
%\end{figure}

\section{Conclusion \label{sec:con}}
Measurements of inclusive jet, dijet mass, $W/Z$ + inclusive jet cross 
sections provide stringent tests of pQCD and are in agreement with NLO 
predictions. The Run II inclusive jet cross section results have 
decreased the central value and uncertainty of gluon PDF at high $x$. 
The dijet mass spectrum and angular distributions have been used to 
set the world's best limits on parameters predicted by 
new physics, such as mass of excited quark, axigluon/coloron, and 
compositeness scale, {\it etc}. Measurement of $b$-jet shape suggests that 
the fraction of gluon-splitting for $b\bar{b}$ production has to be increased 
in {\tt PYTHIA} and {\tt HERWIG}. More data are being collected at the 
Tevatron and 8~\fbarn\ of \ppbar\ collisions are expected by the end of 2010. 
Updates with more data will benefit the $W/Z$ + heavy flavor 
measurements and also push the other analyses to a wider kinematic range. In 
addition, full NLO predictions for $W/Z$ + heavy flavor in a user-friendly MC 
program will give more sensible data and theory comparisons. As the QCD 
productions of these processes are well measured and studied, 
our chance of discovery will be enhanced due to the better understanding of 
backgrounds.

\section{Acknowledgments}
I would like to thank the CDF and D\O\ QCD conveners and the authors 
of each individual analysis, for answering my endless questions.

% ****************************************************************************
% BIBLIOGRAPHY AREA
% ****************************************************************************

\begin{footnotesize}
% IF YOU DO NOT USE BIBTEX, USE THE FOLLOWING SAMPLE SCHEME FOR THE REFERENCES
% ----------------------------------------------------------------------------

\end{footnotesize}

% ****************************************************************************
% END OF BIBLIOGRAPHY AREA
% ****************************************************************************

\end{document}